\RequirePackage{ifpdf}
\ifpdf % We are running pdfTeX in pdf mode
\documentclass[pdftex]{sigma}
\else
\documentclass{sigma}
\fi

\begin{document}

\allowdisplaybreaks

\renewcommand{\thefootnote}{$\star$}

\renewcommand{\PaperNumber}{011}

\FirstPageHeading

\ShortArticleName{On Integrability of a Special Class of Hydrodynamic-Type Systems}

\ArticleName{On Integrability of a Special Class of Two-Component
(2+1)-Dimensional Hydrodynamic-Type Systems\footnote{This paper is a contribution to the Proceedings of the XVIIth International Colloquium on Integrable Systems and Quantum Symmetries (June 19--22, 2008, Prague, Czech Republic). The full collection
is available at
\href{http://www.emis.de/journals/SIGMA/ISQS2008.html}{http://www.emis.de/journals/SIGMA/ISQS2008.html}}}

\Author{Maxim V. PAVLOV~$^\dag$ and Ziemowit POPOWICZ~$^\ddag$}

\AuthorNameForHeading{M.V. Pavlov and Z. Popowicz}

\Address{$^\dag$~Department of Mathematical Physics, P.N.~Lebedev Physical Institute of RAS,\\
 \hphantom{$^\dag$}~53 Leninskii Ave., 119991 Moscow, Russia}
\EmailD{\href{mailto:M.V.Pavlov@lboro.ac.uk}{M.V.Pavlov@lboro.ac.uk}}

\Address{$^\ddag$~Institute of Theoretical Physics, University of Wroc\l aw,\\
 \hphantom{$^\ddag$}~pl.~M.~Borna 9,  50-204 Wroc\l aw, Poland}
\EmailD{\href{mailto:ziemek@ift.uni.wroc.pl}{ziemek@ift.uni.wroc.pl}}

\ArticleDates{Received August 28, 2008, in f\/inal form January 20,
2009; Published online January 27, 2009}

\Abstract{The  particular case of the integrable two component (2+1)-dimensional
hydrodynamical type systems,  which generalises the so-called Hamiltonian
subcase, is considered. The associated system in
involution is integrated in a parametric form. A dispersionless Lax
formulation is found.}

\Keywords{hydrodynamic-type system; dispersionless Lax representation}

\Classification{37K10; 35Q53}

\section{Introduction}

Quasilinear (2+1)-dimensional systems of the f\/irst order,
\begin{equation*}
A_{k}^{i}(\mathbf{u})u_{t}^{k}+B_{k}^{i}(\mathbf{u})u_{y}^{k}+C_{k}^{i}(%
\mathbf{u})u_{x}^{k}=0,
\end{equation*}%
play an important role in the description of variety of physical phenomena.
The method of the hydrodynamical reductions (see e.g.~\cite{Fer+Kar0})
enables us  to pick from this class the integrable systems which
possess suf\/f\/iciently many hydrodynamic reductions, and thus inf\/initely many
particular solutions. Recently, a system in involution describing the integrable
(2+1)-dimensional hydrodynamical type systems%
\begin{equation}
\left(
\begin{array}{c}
v \\
w%
\end{array}%
\right) _{t}=\left(
\begin{array}{cc}
A_{11} & A_{12} \\
A_{21} & A_{22}%
\end{array}%
\right) \left(
\begin{array}{c}
v \\
w%
\end{array}%
\right) _{y}+\left(
\begin{array}{cc}
B_{11} & B_{12} \\
B_{21} & B_{22}%
\end{array}%
\right) \left(
\begin{array}{c}
v \\
w%
\end{array}%
\right) _{x},  \label{!}
\end{equation}%
where $A_{ik}$ and $B_{ik}$ are functions of $v$ and $w$, was derived in
\cite{Fer+Kar} using the method of hydrodynamic reductions. In a particular
case
\begin{equation}
\left(
\begin{array}{c}
v \\
w%
\end{array}%
\right) _{t}=\left(
\begin{array}{cc}
\alpha & 0 \\
0 & \beta%
\end{array}%
\right) \left(
\begin{array}{c}
v \\
w%
\end{array}%
\right) _{y}+\left(
\begin{array}{cc}
p & q \\
r & s%
\end{array}%
\right) \left(
\begin{array}{c}
v \\
w%
\end{array}%
\right) _{x},  \label{1}
\end{equation}%
where $\alpha $ and $\beta $ are constants, the corresponding system in
involution for two functions $r(v,w)$ and $q(v,w)$ simplif\/ies to the form%
\begin{gather}
q_{vv}  = (qr)_{w},\qquad q_{vw}=\frac{q_{v}q_{w}}{q}+q^{2}r, \qquad q_{ww}=\frac{q_{w}^{2}}{q}+2qq_{v}-\frac{q_{w}r_{w}}{r},  \notag \\
   \label{2}
r_{ww}  = (qr)_{v},\qquad r_{vw}=\frac{r_{v}r_{w}}{r}+r^{2}q, \qquad r_{vv}=\frac{r_{v}^{2}}{r}+2rr_{w}-\frac{q_{v}r_{v}}{q},
\end{gather}%
and a general solution therefore  depends on 6 arbitrary constants. Two other
functions $p(v,w)$ and $s(v,w)$ can be found in quadratures%
\begin{equation}
dp=rqdv+q_{v}dw, \qquad ds=r_{w}dv+rqdw.  \label{4}
\end{equation}%
A general solution of system (\ref{2}) was presented in \cite{Fer+Kar} in a
parametric form.

Under a simple linear transformation of independent variables ($y,t$) such
that%
\begin{equation*}
\partial _{t}-\alpha \partial _{y}\rightarrow \partial _{t},\qquad
\partial _{t}-\beta \partial _{y}\rightarrow \partial _{y},
\end{equation*}%
system (\ref{1}) reduces to the form%
\begin{equation}
v_{t}=p(v,w)v_{x}+q(v,w)w_{x},\qquad w_{y}=r(v,w)v_{x}+s(v,w)w_{x}.
\label{5}
\end{equation}%
Our f\/irst result is that the equations (\ref{5}) reduces to the most compact form%
\begin{equation}
v_{t}=\partial _{x}H_{ww},\qquad w_{y}=\partial _{x}H_{vv}  \label{new}
\end{equation}%
by introducing the potential function $H(v,w)$ such that $r=H_{vvv}$,  $s=H_{vvw}$,   $p=H_{vww}$,   $q=H_{www}$. Then the
equations (\ref{2}) can  be written in the form%
\begin{gather}
H_{vvww}  =  H_{vvv}H_{www},  \notag \\
H_{vvvvw}  = \frac{H_{vvvv}H_{vvvw}}{H_{vvv}}+H_{vvv}^{2}H_{www},
\notag \\
H_{vwwww}  = \frac{H_{wwww}H_{vwww}}{H_{www}}+H_{www}^{2}H_{vvv},
 \notag\\
H_{vvvvv} = \frac{H_{vvvv}^{2}}{H_{vvv}}-\frac{H_{vvvv}H_{vwww}}{H_{www}}%
+2H_{vvv}H_{vvvw},  \notag \\
H_{wwwww}  = \frac{H_{wwww}^{2}}{H_{www}}-\frac{H_{wwww}H_{vvvw}}{H_{vvv}}%
+2H_{www}H_{vwww}. \label{five}
\end{gather}%
Our second result is that we present a general solution of system in
involution (\ref{five}) in a new parametric form (\ref{z}), which is more
convenient for the investigation in the  special cases. This approach is universal,
i.e.\ any other (2+1)-dimensional quasilinear equations
can be investigated ef\/fectively in the same fashion (see~\cite{ops,  odes}).

In the general case (\ref{!}), a dispersionless Lax representation has the
form (see \cite{Fer+Kar})%
\begin{equation}
\psi _{t}=a(\psi _{x},v,w),\qquad \psi _{y}=c(\psi _{x},v,w),
\label{7}
\end{equation}%
where $a(\mu ,v,w)$ and $c(\mu ,v,w)$ are some functions. The function $\psi$
is called a pseudopotential. Equations (\ref{2}) and (\ref{4}) can be
derived directly from the compatibility condition $(\psi _{t})_{y}=(\psi
_{y})_{t}$.

Our third result is that the dispersionless Lax representation (\ref{7}) for
(2+1)-dimensional hyd\-rodynamical type system (\ref{5}) reduces to the most
compact form%
\begin{equation}
\psi _{t}=a(\psi _{x},w),\qquad \psi _{y}=c(\psi _{x},v),
\label{lax}
\end{equation}%
i.e., each of the functions $a$ and $c$ depends on two arguments only.

Moreover, we were able to integrate the corresponding nonlinear system for
the functions $a(\mu ,w)$ and $c(\mu ,v)$, i.e.\ we found the dispersionless
Lax pair (\ref{lax}) for the (2+1)-dimensional quasilinear equations (\ref{new}). In general, the compatibility condition $(\psi _{t})_{y}=(\psi _{y})_{t}$
for the system
\begin{equation*}
\psi_t=a,\qquad \psi_y=c,
\end{equation*}
where $a=a(\mu ,u^{1},u^{2},\dots ,u^{N})$ and $c=c(\mu ,u^{1},u^{2},\dots ,u^{N})$
are some functions (cf.~(\ref{lax})), yields (2+1)-dimensional quasilinear
systems of the nonlinear equations of the f\/irst order.
Recently, a complete classif\/ication of pseudopotentials that satisfy
the single equation $\psi _{t}=a(\psi _{x},w)$ was given in~\cite{ops}.

Our fourth result is that we were able to extract the functions $a(\mu ,w)$
and $c(\mu ,v)$ such that~(\ref{lax}) yields a system of the form~(\ref{new}).

Let us mention that the particular case, when $H_{vvv}=H_{www}$,
\begin{equation}
v_{t}=\partial _{x}h_{v},\qquad w_{y}=\partial _{x}h_{w}.  \label{6}
\end{equation}%
was considered in \cite{Fer+Kar0} and a complete classif\/ication of all admissible
functions $h(v,w)$ was presented in~\cite{FMS}.

The paper is organized as follows. In Section~\ref{section2}, we prove that quasilinear
system~(\ref{new}) admits the dispersionless Lax representation (\ref{lax}),
which is a  special case of~(\ref{7}), and conversely, a dispersionless
Lax representation~(\ref{lax}) yields an integrable (2+1)-dimensional
quasilinear system~(\ref{new}). We also derive the associated system in
involution (\ref{2}) which in the case under study takes the form (\ref{five}). The problem of computation of a single function $H(v,w)$ that solves (\ref{five}) is  reduced  to quadratures. In Section~\ref{section3}, we present an
ef\/fective method for direct integration of the system~(\ref{2}) and the
corresponding equations are integrated in the parametric form. The
last section contains the conclusion.

\section{System in involution}\label{section2}

A classif\/ication of integrable (2+1)-dimensional two-component Hamiltonian
hydrodynamic-type systems (\ref{6}) was obtained in \cite{Fer+Kar0} using
the method of hydrodynamic reductions. In this paper, we use dispersionless
Lax representations following the original papers \cite{Fer+Kar} and \cite{zakh+multi} (see~\cite{odes} for further details). It was proved in \cite{Fer+Kar} that the pseudopotentials $\psi (x,y,t)$ for (2+1)-dimensional
hydrodynamic-type systems (\ref{!}) must satisfy the dispersionless Lax
representation of the form~(\ref{7}). Also, it was proved in \cite{FMS} that
there exist the dispersionless Lax representations for (\ref{6}) of the form~(\ref{lax}). Moreover, in our case the dispersionless Lax representation~(\ref{lax}) remains valid for more general case~(\ref{5}). Indeed, the
following assertion holds.

\begin{lemma} The hydrodynamic-type system \eqref{5} can
be obtained from the compatibility condition  $(\psi _{t})_{y}=(\psi
_{y})_{t}$,  where the pseudopotential  $\psi $  satisfies
dispersionless Lax representation \eqref{lax}.
\end{lemma}

\begin{proof} The compatibility condition for %general case
(\ref{7}) implies
\begin{gather*}
\partial _{y}a(\psi _{x},v,w)=\partial _{t}c(\psi _{x},v,w).
\end{gather*}%
Substituting (\ref{5}) into this equation yields
\begin{gather*}
 a_{\mu }(c_{v}v_{x}+c_{w}w_{x})+a_{v}v_{y}+a_{w}(rv_{x}+sw_{x}) \\
\qquad {}= c_{\mu }(a_{v}v_{x}+a_{w}w_{x})+c_{v}(pv_{x}+qw_{x})+c_{w}w_{t},
\end{gather*}%
where $\mu =\psi _{x}$. Since we are interested in a general solution, the
coef\/f\/icients at the $t$- and $x$-derivatives $v$ and $w$ on the left- and
right-hand sides of the above equations must be identical, that is,
\begin{gather*}
a_{\mu }c_{v}+ra_{w}=c_{\mu }a_{v}+pc_{v},\qquad a_{\mu
}c_{w}+sa_{w}=c_{\mu }a_{w}+qc_{v},
\end{gather*}%
and $a_{v}=0,c_{w}=0$. Thus, $a$ is independent of $v$, and $c$ is
independent of $w$. This means that~(\ref{7}) reduces to (\ref{lax}). The
lemma is proved.
\end{proof}

\begin{theorem}   The dispersionless Lax representation \eqref{lax}
 uniquely determines a hydrodynamic-type system of the form \eqref{5}.
 \end{theorem}

\begin{proof} The compatibility condition $(\psi _{t})_{y}=(\psi _{y})_{t}$
implies
\begin{equation}
a_{\mu }c_{v}v_{x}+a_{w}w_{y}=c_{\mu }a_{w}w_{x}+c_{v}v_{t},  \label{comp}
\end{equation}%
where the dispersionless Lax representation (\ref{lax}) is re-written in the
form%
\begin{equation}
\mu _{t}=\partial _{x}a(\mu ,w),\qquad \mu _{y}=\partial
_{x}c(\mu ,v)  \label{gen}
\end{equation}%
Since $a_{w}\neq 0$, dif\/ferentiating the equation (\ref{comp})%
\begin{equation*}
\frac{a_{\mu }c_{v}}{a_{w}}v_{x}+w_{y}=c_{\mu }w_{x}+\frac{c_{v}}{a_{w}}v_{t}
\end{equation*}%
with respect to $\mu$ yields %leads to the equation%
\begin{equation*}
\left( \frac{a_{\mu }c_{v}}{a_{w}}\right) _{\mu }v_{x}=c_{\mu \mu
}w_{x}+\left( \frac{c_{v}}{a_{w}}\right) _{\mu }v_{t}.
\end{equation*}%
Suppose that $(c_{v}/a_{w})_{\mu }\neq 0$, then the equation
\begin{equation}  \label{vt}
v_{t}=\frac{\left( \frac{a_{\mu }c_{v}}{a_{w}}\right) _{\mu }}{\left( \frac{%
c_{v}}{a_{w}}\right) _{\mu }}v_{x}-\frac{c_{\mu \mu }}{\left( \frac{c_{v}}{%
a_{w}}\right) _{\mu }}w_{x}
\end{equation}%
cannot depend on $\mu $. This means that (see (\ref{5})) we have
\begin{equation*}
\frac{\left( \frac{a_{\mu }c_{v}}{a_{w}}\right) _{\mu }}{\left( \frac{c_{v}}{%
a_{w}}\right) _{\mu }}=p(v,w), \qquad \frac{c_{\mu \mu }}{\left(
\frac{c_{v}}{a_{w}}\right) _{\mu }}=-q(v,w)
\end{equation*}%
for some functions $p(v,w)$ and $q(v,w)$ that do not depend on $\mu$. Then
the above equation (\ref{vt}) reduces to the form (cf.~(\ref{5}))%
\begin{equation*}
v_{t}=p v_{x}+qw_{x}.
\end{equation*}%
Since $c_{v}\neq 0$, two other relations of the above type,
\begin{equation*}
\frac{a_{\mu \mu }}{\left( \frac{a_{w}}{c_{v}}\right) _{\mu }}=-r(v,w),\qquad \frac{\left( \frac{c_{\mu }a_{w}}{c_{v}}\right) _{\mu }}{\left(
\frac{a_{w}}{c_{v}}\right) _{\mu }}=s(v,w),
\end{equation*}%
where $r(v,w)$ and $s(v,w)$ are some other functions independent of $\mu$,
can be derived in analogy with the above, along with the second equation of
the hydrodynamic-type system~(\ref{5}). Thus, the hydrodynamic-type system~(\ref{5}) is indeed uniquely determined by the compatibility condition $(\psi
_{t})_{y}=(\psi _{y})_{t}$. The theorem is proved.
\end{proof}

Substituting (\ref{5}) into (\ref{comp}) implies the relations%
\begin{equation}
a_{w}=\frac{qc_{v}}{s-c_{\mu }},\qquad rq=(s-c_{\mu })(p-a_{\mu }).
\label{15}
\end{equation}%
among the f\/irst derivatives of the functions $a(\mu ,w)$ and $c(\mu ,v)$.

The compatibility conditions $(a_{w})_{\mu }=(a_{\mu })_{w}$,   $%
(a_{w})_{v}=0$ and $(a_{\mu })_{v}=0$ yield%
\begin{gather}
c_{vv}  = \frac{p_{v}}{rq}c_{\mu }c_{v}+\frac{qr_{v}-sp_{v}}{rq}c_{v},
\notag \\
c_{\mu v}  = \frac{p_{v}}{rq}c_{\mu }^{2}+\left( \frac{q_{v}}{q}+\frac{r_{v}%
}{r}-2\frac{sp_{v}}{rq}\right) c_{\mu }+\frac{s^{2}p_{v}}{rq}-\left( \frac{%
q_{v}}{q}+\frac{r_{v}}{r}\right) s+s_{v},  \label{j} \\
c_{\mu \mu }  = \frac{1}{qc_{v}}\left[ \frac{p_{v}}{r}(c_{\mu
}-s)^{3}+\left( p_{w}+q_{v}+\frac{qr_{v}}{r}\right) (c_{\mu
}-s)^{2}+(rq_{w}+qr_{w}+qs_{v})(c_{\mu }-s)+rqs_{w}\right] .  \notag
\end{gather}%
Similar formulas can be obtained from the other compatibility conditions,
namely, $(c_{v})_{\mu }=(c_{\mu })_{v}$,   $(c_{v})_{w}=0$ and $(c_{\mu
})_{w}=0$. The compatibility conditions $(c_{\mu \mu })_{v}=(c_{\mu v})_{\mu
}$,   $(c_{\mu v})_{v}=(c_{vv})_{\mu }$,   $(a_{\mu \mu })_{w}=(a_{\mu
w})_{\mu }$,   $(a_{\mu w})_{w}=(a_{ww})_{\mu }$ yield the system%
\begin{gather}
p_{vv}  = \frac{p_{v}}{rq}(rp_{w}+qr_{v}),\qquad p_{vw}=\frac{p_{v}%
}{rq}(rq_{w}+qr_{w}),\qquad p_{ww}=\frac{1}{rq}%
(rp_{w}q_{w}+qp_{v}s_{w}), \nonumber\\
s_{ww}  = \frac{s_{w}}{rq}(rq_{w}+qs_{v}), \qquad s_{vw}=\frac{s_{w}%
}{rq}(rq_{v}+qr_{v}),\qquad s_{vv}=\frac{1}{rq}%
(qr_{v}s_{v}+rp_{v}s_{w}),\nonumber
\\
rq_{vv}+qr_{vv}  = \frac{1}{rq}%
(r^{2}p_{v}q_{w}+q^{2}r_{v}^{2})+2s_{v}p_{v}+r_{w}p_{v}-r_{v}q_{v},
\label{com} \\
rq_{ww}+qr_{ww}  = \frac{1}{rq}%
(q^{2}s_{w}r_{v}+r^{2}q_{w}^{2})+2p_{w}s_{w}+q_{v}s_{w}-q_{w}r_{w},  \notag \\
rq_{vw}+qr_{vw}  = \frac{1}{rq}(r^{2}q_{v}q_{w}+q^{2}r_{v}r_{w})+2p_{v}s_{w}.
\notag
\end{gather}%
The last equation can be replaced by the pair of equations%
\begin{gather*}
r_{vw}=\frac{1}{rq}(rp_{v}s_{w}+qr_{v}r_{w}),\qquad q_{vw}=\frac{1}{rq%
}(rq_{v}q_{w}+qp_{v}s_{w}),
\end{gather*}%
which can be obtained from the compatibility conditions $%
(p_{vw})_{v}=(p_{vv})_{w}$ and $(s_{vw})_{w}=(s_{ww})_{v}$. The second and
f\/ifth equations of (\ref{com}) can be integrated once to yield%
\begin{equation}
p_{v}=rq\varphi _{1}(v),\qquad s_{w}=rq\varphi _{2}(w),
\label{choice}
\end{equation}%
where $\varphi _{1}(v)$ and $\varphi _{2}(w)$ are arbitrary functions.
However, without loss of generality these functions can be set equal to 1,
because these functions can be eliminated from all of the above equations
upon using the scaling $\int \varphi _{1}(v)dv\rightarrow v$,   $\int
\varphi _{2}(w)dw\rightarrow w$,   $\varphi _{1}(v)p/\varphi
_{2}(w)\rightarrow q$,  $\varphi _{2}(w)r/\varphi _{1}(v)\rightarrow r$.
Thus, substituting $p_{v}=rq$,  $s_{w}=rq$ into (\ref{com}), we f\/inally
obtain the system in involution (\ref{2}) together with (\ref{4}) (cf.~\cite{Fer+Kar}).

Thus, integrable (2+1)-dimensional system (\ref{5}) can be written in the
most compact form (\ref{new}), and the reduction to the Hamiltonian case (\ref{6}) is given by the symmetric constraint $r=q$. The function $H$ can be
reconstructed via the complete dif\/ferentials,%
\begin{alignat*}{4}
& dH_{vv}  = rdv+sdw,\qquad && dH_{vw}=sdv+pdw,\qquad && dH_{ww}=pdv+qdw, &  \\
& dH_{v}  = H_{vv}dv+H_{vw}dw,\qquad && dH_{w}=H_{vw}dv+H_{ww}dw,\qquad && dH=H_{v}dv+H_{w}dw.&
\end{alignat*}

\begin{remark} The above choice (\ref{choice}) uniquely f\/ixes system (\ref{5}) in conservative form (\ref{new}). This means that the integrable
(2+1)-dimensional hydrodynamic-type system (\ref{5}) is reducible to (\ref%
{new}) under appropriate choice of functions $\varphi _{1}(v)$ and $\varphi
_{2}(w)$ in (\ref{choice}).
\end{remark}

\section{Dispersionless Lax representation}\label{section3}

In this section, we obtain a general solution of the system in involution (\ref{five}) and simultaneously reconstruct the functions $a(\mu ,w)$ and $c(\mu ,v)$ (see (\ref{lax})).

Rewrite system (\ref{j}) for the function $c(\mu ,v)$ in the form%
\begin{gather}
c_{vv}  = A_{1}c_{\mu }c_{v}+A_{2}c_{v},  \notag \\
c_{\mu v}  = B_{1}c_{\mu }^{2}+B_{2}c_{\mu }+B_{3},  \label{si} \\
c_{v}c_{\mu \mu }  = D_{1}c_{\mu }^{3}+D_{2}c_{\mu }^{2}+D_{3}c_{\mu }+D_{4},
\notag
\end{gather}%
where all the coef\/f\/icients $A_{k}$, $B_{n}$, $D_{m}$ depend on $v$ alone and are
to be determined. The compatibility conditions $(c_{vv})_{\mu }=(c_{\mu
v})_{v},(c_{\mu \mu })_{v}=(c_{\mu v})_{\mu }$ give rise to the following
system for the coef\/f\/icients $A_{k}$, $B_{n}$, $D_{m}$:
\begin{gather}
A_{1}  = B_{1}=D_{1},  \notag \\
B_{1}^{\prime } =A_{2}B_{1}-2B_{1}B_{2}+B_{1}D_{2},  \notag \\
B_{2}^{\prime } =A_{2}B_{2}-B_{1}B_{3}+B_{1}D_{3}-B_{2}^{2},  \notag \\
B_{3}^{\prime } =A_{2}B_{3}+B_{1}D_{4}-B_{2}B_{3},  \label{x} \\
D_{2}^{\prime } = A_{2}D_{2}-3B_{1}B_{3}+2B_{1}D_{3}-B_{2}D_{2},  \notag \\
D_{3}^{\prime } = A_{2}D_{3}+3B_{1}D_{4}-2B_{3}D_{2},  \notag \\
D_{4}^{\prime }  = A_{2}D_{4}+B_{2}D_{4}-B_{3}D_{3},  \notag
\end{gather}
where the prime denotes the derivative with respect to $v$.

The derivative $c_{v}$ can be expressed from the last equation of (\ref{si}). Plugging this expression into the l.h.s.\ of the second equation of (\ref{si}) yields
\begin{equation}
\frac{c_{\mu \mu \mu }}{c_{\mu \mu }^{2}}=\frac{(3D_{1}-B_{1})c_{\mu
}^{2}+(2D_{2}-B_{2})c_{\mu }+D_{3}-B_{3}}{D_{1}c_{\mu }^{3}+D_{2}c_{\mu
}^{2}+D_{3}c_{\mu }+D_{4}},  \label{8}
\end{equation}%
which can be expanded into simple fractions (see \cite{ops} for the general
case)
\begin{equation}
\frac{c_{\mu \mu \mu }}{c_{\mu \mu }^{2}}=\frac{k_{1}}{c_{\mu }-b_{1}}+
\frac{k_{2}}{c_{\mu }-b_{2}}+\frac{k_{3}}{c_{\mu }-b_{3}},  \label{9}
\end{equation}%
where $k_{i}(v)$ are some functions such that $k_{1}+k_{2}+k_{3}=2$, and the
functions $b_{k}(v)$ are roots of the cubic polynomial in the last equation
of (\ref{si}):
\begin{equation*}
c_{v}c_{\mu \mu }=D_{1}(c_{\mu }-b_{1})(c_{\mu }-b_{2})(c_{\mu }-b_{3}).
\end{equation*}%
Conversely, upon comparing (\ref{8}) with (\ref{9}) the above coef\/f\/icients $%
A_{k}$, $B_{n}$, $D_{m}$ can be expressed via the functions $b_{k}(v)$ in the
symmetric form
\begin{gather*}
-B_{2}/D_{1}  = k_{1}b_{1}+k_{2}b_{2}+k_{3}b_{3},\qquad
B_{3}/D_{1}=(1-k_{1})b_{2}b_{3}+(1-k_{2})b_{1}b_{3}+(1-k_{3})b_{1}b_{2}, \\
-D_{2}/D_{1}  = b_{1}+b_{2}+b_{3},\qquad
D_{3}/D_{1}=b_{1}b_{2}+b_{1}b_{3}+b_{2}b_{3},\qquad
-D_{4}/D_{1}=b_{1}b_{2}b_{3}.
\end{gather*}

\begin{lemma} $k_{i}$  are constants.
\end{lemma}

\begin{proof} Integrating (\ref{9}) yields the following equation:
\begin{equation*}
c_{\mu \mu }=b(c_{\mu }-b_{1})^{k_{1}}(c_{\mu }-b_{2})^{k_{2}}(c_{\mu
}-b_{3})^{k_{3}},
\end{equation*}%
where $b(v)$ is a function of $v$ alone. Then (see the last equation in (\ref{si})) we have
\begin{equation}
c_{v}=\frac{D_{1}}{b}(c_{\mu }-b_{1})^{1-k_{1}}(c_{\mu
}-b_{2})^{1-k_{2}}(c_{\mu }-b_{3})^{1-k_{3}}.  \label{17}
\end{equation}%
The compatibility condition $(c_{v})_{\mu \mu }=(c_{\mu \mu })_{v}$ is
satisf\/ied only if $k_{i}$ are constants. The lemma is proved.
\end{proof}

\begin{theorem} Under the above substitutions, the system \eqref{x} reduces to the form
\begin{equation}
b_{i}^{\prime }=D_{1}(1-k_{i})\prod_{j\neq i}(b_{i}-b_{j}),\qquad i=1,2,3,
\label{tri}
\end{equation}%
In this case we have
\begin{equation}
A_{2}=D_{1}^{\prime
}/D_{1}+D_{1}(b_{1}+b_{2}+b_{3})-2D_{1}(k_{1}b_{1}+k_{2}b_{2}+k_{3}b_{3}).
\label{fo}
\end{equation}
\end{theorem}

\begin{proof} The coef\/f\/icient $A_{2}$ in the form (\ref{fo}) can be
expressed from the second equation of (\ref{x}). The last three equations of
(\ref{x}) are linear with respect to the f\/irst derivatives $b_{k}^{\prime }$. Solving this linear system with respect to $b_k$ immediately yields (\ref{tri}). Moreover, the third and fourth equations of (\ref{x}) are then
automatically satisf\/ied. The theorem is proved.
\end{proof}

Let us choose $A_{1}=B_{1}=D_{1}=1$ in this formulation (see formulae (\ref{si}), (\ref{x}), (\ref{tri}), (\ref{fo})) in agreement with the
normalization (\ref{choice}). This means that a solution of the system (see (%
\ref{tri}), where $D_{1}=1$)%
\begin{equation}
b_{i}^{\prime }=(1-k_{i})\prod_{j\neq i}(b_{i}-b_{j}),\qquad i=1,2,3
\label{3}
\end{equation}%
determines the coef\/f\/icients $p$, $q$, $r$, $s$ of (2+1)-dimensional integrable
hydrodynamic-type system~(\ref{5}) written in the form~(\ref{new}).

Introducing  the  ``intermediate'' independent variable $V(v)$ such that%
\begin{equation}
V^{\prime }=\xi V^{k_{1}}(1-V)^{k_{3}},  \label{subs}
\end{equation}%
where $\xi $ is an arbitrary constant we obtain the following theorem:

\begin{theorem} General solution of system \eqref{3}  can
be written in the form
\begin{gather*}
b_{2}  = b_{1}+\xi V^{k_{1}-1}(1-V)^{k_{3}},\qquad
b_{3}=b_{1}+\xi V^{k_{1}-1}(1-V)^{k_{3}-1}, \\
b_{1}  = (1-k_{1})\xi \int V^{k_{1}-2}(1-V)^{k_{3}-1}dV.
\end{gather*}
\end{theorem}

\begin{proof} Introduce the auxiliary functions $b_{12}=b_{2}-b_{1}$ and $%
b_{13}=b_{3}-b_{1}$. Then the system (\ref{3}) reduces to the form%
\begin{gather}
b_{1}^{\prime }=(1-k_{1})b_{12}b_{13},\qquad b_{12}^{\prime
}=b_{12}[(1-k_{2})b_{12}-k_{3}b_{13}],\nonumber\\
 b_{13}^{\prime
}=b_{13}[(1-k_{3})b_{13}-k_{2}b_{12}].  \label{bs}
\end{gather}%
The ratio of the last two equations%
\begin{equation*}
\frac{d\ln b_{12}}{d\ln b_{13}}=\frac{(1-k_{2})b_{12}-k_{3}b_{13}}{%
(1-k_{3})b_{13}-k_{2}b_{12}}
\end{equation*}%
is nothing but a f\/irst-order ODE. Substituting the intermediate function $%
V=1-b_{12}/b_{13}$ into this ODE reduces to the following
quadrature:
\begin{equation*}
d\ln b_{13}=(k_{3}-1)d\ln (1-V)+(k_{1}-1)d\ln V.
\end{equation*}%
Taking into account that $b_{12}=(1-V)b_{13}$, one can obtain the equality $%
b_{12}=\xi V^{k_{1}-1}(1-V)^{k_{3}}$, where the above quadrature is
integrated in the parametric form $b_{13}=\xi V^{k_{1}-1}(1-V)^{k_{3}-1}$.
Substituting these expressions for $b_{12}$ and $b_{13}$ into the f\/irst
equation of (\ref{bs}), $b_{1}^{\prime }=(1-k_{1})b_{12}b_{13}$, yields the
quadrature%
\begin{equation*}
db_{1}=(1-k_{1})\xi V^{k_{1}-2}(1-V)^{k_{3}-1}dV.
\end{equation*}%
The remaining equations in (\ref{bs}) yield (\ref{subs}). The theorem is
proved.
\end{proof}

In turn, comparing the expressions for $c_{\mu\mu}$, $c_{\mu v}$, and $c_{vv}$
from~(\ref{j}) with
their counter\-parts~(\ref{si}) and equating the coef\/f\/icients at the powers of
$c_\mu$ and $c_v$
gives rise the following quadratures:
\begin{gather}
dp  = q[rdv+(D_{2}-B_{2}+s)dw],  \label{13} \\
d\ln q  = \frac{2s^{2}+(2D_{2}-B_{2})s+D_{3}-B_{3}}{r}dw+(B_{2}+2s)dv-d\ln r,
\label{12}
\end{gather}%
where%
\begin{equation}
r=\frac{s^{3}+D_{2}s^{2}+D_{3}s+D_{4}}{s_{w}},  \label{14}
\end{equation}%
and the function $s(v,w)$ satisf\/ies the Riccati equation
\begin{equation}
s_{v}=s^{2}+B_{2}s+B_{3}.  \label{sik}
\end{equation}

This equation can be reduced to the linear ODE%
\begin{equation*}
f_{v}=[k_{3}(b_{3}-b_{1})+k_{2}(b_{2}-b_{1})]f-1
\end{equation*}%
by the substitution $s=b_{1}+1/f(v,w)$ with the  solution
given by
\begin{equation*}
f=\frac{1-VW}{b_{2}-b_{1}},
\end{equation*}%
where $W(w)$ is an integration ``constant''. With this in mind, the function $r$ can be found from (\ref{14}). In turn,
the function $W(w)$ cannot be found from the compatibility conditions $%
\partial _{v}(\partial _{w}\ln q)=\partial _{w}(\partial _{v}\ln q)$ (see~(\ref{12})) or $\partial _{v}(p_{w})=\partial _{w}(p_{v})$ (see~(\ref{13})).
A substitution of $q=s_{w}/r$ (see (\ref{4})) into (\ref{12}) yields an
equation
\begin{equation}
W^{\prime }=\bar{\xi}W^{k_{2}}\left( 1-W\right) ^{k_{3}}  \label{pod}
\end{equation}%
which is similar to (\ref{subs}). Here $\bar{\xi}$ is an arbitrary constant.
Then the two quadratures~(\ref{13}) and~(\ref{12}) can be performed
explicitly.

Thus, the functions $p$, $q$, $r$, $s$ are given by the following expressions:
\begin{gather}
s  = \xi \frac{V^{k_{1}-1}(1-V)^{k_{3}}}{1-VW}+(1-k_{1})\xi \int
V^{k_{1}-2}(1-V)^{k_{3}-1}dV,  \notag \\
r  = -\frac{\xi ^{2}}{\bar{\xi}}\frac{W^{1-k_{2}}\left( 1-W\right) ^{1-k_{3}}%
}{1-VW}V^{2k_{1}-1}(1-V)^{2k_{3}-1},  \notag \\
q =-\frac{\bar{\xi}^{2}}{\xi }W^{2k_{2}-1}\left( 1-W\right) ^{2k_{3}-1}%
\frac{V^{1-k_{1}}(1-V)^{1-k_{3}}}{1-VW},  \notag \\
p =\bar{\xi}\frac{W^{k_{2}-1}\left( 1-W\right) ^{k_{3}}}{1-VW}+(1-k_{2})%
\bar{\xi}\int W^{k_{2}-2}(1-W)^{k_{3}-1}dW.  \label{z}
\end{gather}%
This means that the function $H(v,w)$ (see (\ref{five})) is determined via
its third derivatives (see the end of Section~\ref{section2}). Thus, we proved that a
single function $c(\mu ,v)$ completely determines a (2+1)-dimensional
quasilinear system~(\ref{new}), and the second function $a(\mu ,w)$ is
determined via the f\/irst derivatives of~$c$:
\begin{equation}
da=\frac{qc_{v}}{s-c_{\mu }}dw+\left( p-\frac{rq}{s-c_{\mu }}\right) d\mu .
\label{t}
\end{equation}

In  order to compute the function $c(\mu ,v)$ we integrate  the second
equation in (\ref{si}). Indeed, the equation in question (recall that we
have $B_{1}=1$ in this normalization)
\begin{equation*}
\partial _{v}c_{\mu }=c_{\mu }^{2}+B_{2}c_{\mu }+B_{3}
\end{equation*}%
coincides with (\ref{sik}) up to the replacement $c_{\mu }\leftrightarrow s$. Since $D_{1}=1$, the compatibility condition $(c_{v})_{\mu \mu }=(c_{\mu
\mu })_{v} $ implies $b(v)=1$ in~(\ref{17}). Thus, the derivative $c_{v}$
can also be found. Finally, substituting (\ref{z}) and just obtained expressions for $c_{\mu }$, $c_{v}$ into (\ref{t}) yields the corresponding dispersionless Lax
representation (see (\ref{gen}), (\ref{15}), (\ref{si})) which is now
determined by means of the formulas
\begin{gather*}
c_{\mu }  = \xi \frac{V^{k_{1}-1}(1-V)^{k_{3}}}{1-\epsilon V}+(1-k_{1})\xi
\int V^{k_{1}-2}(1-V)^{k_{3}-1}dV,  \notag \\
c_{V}  = -\frac{\xi }{\tilde{\xi}}\epsilon ^{1-k_{2}}(1-\epsilon )^{1-k_{3}}%
\frac{V^{k_{1}-1}(1-V)^{k_{3}-1}}{1-\epsilon V},  \notag \\
a_{W}  = -\frac{\tilde{\xi}}{\xi }\epsilon ^{1-k_{1}}(1-\epsilon )^{1-k_{3}}%
\frac{W^{k_{2}-1}(1-W)^{k_{3}-1}}{1-\epsilon W},  \notag \\
a_{\mu }  = \tilde{\xi}\frac{W^{k_{2}-1}(1-W)^{k_{3}}}{1-\epsilon W}%
+(1-k_{2})\tilde{\xi}\int W^{k_{2}-2}(1-W)^{k_{3}-1}dW, % \label{y}
\end{gather*}%
where the auxiliary variable $\epsilon(\mu)$ is determined by the formula
(cf.~(\ref{pod}))%
\begin{equation*}
\epsilon ^{\prime }(\mu )=\xi _{0}\epsilon ^{k_{2}}(1-\epsilon )^{k_{3}}.
\end{equation*}

\section{Conclusion}\label{section4}

As it was mentioned in the Introduction, the integrable (2+1)-dimensional
quasilinear systems of the nonlinear f\/irst-order equations (see \cite{odes} for
details) are determined by the compatibility condition $(\psi_{t})_{y}=(\psi_{y})_{t}$ , where in general $\psi_t=a(\mu ,u^{1},u^{2},\dots ,u^{N})$ and
$\psi_y=c(\mu ,u^{1},u^{2},\dots ,u^{N})$, cf.~(\ref{7}) and (\ref{lax}). An
open problem is whether it is possible to construct hydrodynamic chains
(see, for instance, \cite{Fer+Mar}) associated with such (2+1)-dimensional
quasilinear systems. The theory of integrable hydrodynamic chains is much
simpler than the theory of integrable (2+1)-dimensional quasilinear
equations, because the former still is a theory of integrable
(1+1)-dimensional hydrodynamic-type systems with just one  nontrivial
extension -- allowing for inf\/initely many components. Thus, an integrable
hydrodynamic chain possesses the properties that are well-known in the
theory of f\/inite-component systems (dispersive or dispersionless), such as
inf\/inite series of conservation laws, inf\/inite series of commuting f\/lows,
and inf\/inite series of Hamiltonian structures.

At least, we can answer the above question regarding the construction of the
associated hydrodynamic chain for the case of~(\ref{lax}), but this will be
the subject of a separate paper. Moreover, if we f\/ix the f\/irst equation in (\ref{lax}), then an associated integrable hierarchy can be found from the
dispersionless Lax representations (cf.~(\ref{7}) and~(\ref{lax}))%
\begin{equation*}
\psi _{t}=a(\psi _{x},w),\qquad \psi _{y^{N}}=c\big(\psi
_{x},v^{1},v^{2},\dots ,v^{N}\big),
\end{equation*}%
where the f\/irst member of this hierarchy is given by (\ref{new}) and
uniquely determined by the dispersionless Lax representation (\ref{lax}).
This means that inf\/initely many commuting f\/lows (numbered by the
``times'' $y^{k}$) will be determined. It
would be interesting to f\/ind the associated hydrodynamic chains for the case
when, instead of the above dispersionless Lax representation, one has a more
general ansatz%
\begin{equation*}
\psi _{t}=a\big(\psi _{x},w^{1},w^{2},\dots ,w^{M}\big),\qquad \psi
_{y^{N}}=c\big(\psi _{x},v^{1},v^{2},\dots ,v^{N}\big),
\end{equation*}%
where $M$ and $N$ are arbitrary positive integers.

\subsection*{Acknowledgements}

We thank Eugeni Ferapontov, Sergey Tsarev and Sergey Zykov for their
stimulating and clarifying discussions.
M.V.P.~would like to thank the Institute of Theoretical Physics of Wroc\l aw
University for the hospitality and the Kasa Mianowski Foundation for the
f\/inancial support of MVP's visit to Wroc\l aw making this collaboration
possible.
MVP is grateful to professor Boris Dubrovin for a hospitality in SISSA in
Trieste (Italy) where part of this work has been done. MVP was partially
supported by the Russian-Italian Research Project (Consortium E.I.N.S.T.E.IN
and RFBR grant 06-01-92053).

\pdfbookmark[1]{References}{ref}
\LastPageEnding

\end{document}